# Quantum Software Models: The Density Matrix for Classical and Quantum Software Systems Design

Iaakov Exman
Software Engineering Department
The Jerusalem College of Engineering – Azrieli
Jerusalem, Israel
iaakov@jce.ac.il

Alon Tsalik Shmilovich
Software Engineering Department
The Jerusalem College of Engineering – Azrieli
Jerusalem, Israel
alonshmilo@gmail.com

*Abstract*— Linear Software Models enable rigorous linear algebraic procedures for modular design of classical software systems. These procedures apply a spectral approach to matrix representations – e.g. the Laplacian – of the software system.

Recent intensive research efforts towards quantum computers have increased expectations that quantum computing could in due time materialize as a practical alternative to classical computing. It is reasonable to inquire about quantum software desirable features and prepare in advance modular design procedures for quantum software systems.

However, it does not make sense to have two totally separate procedures for modular design, one for classical software systems and another for quantum software systems. This paper claims that there should be just a single unified and rigorous design procedure for both classical and quantum software systems.

Our common design procedure starting point for both classical and quantum software systems is Von Neumann's quantum notion of Density Operator and its Density Matrix representation. This paper formulates and demonstrates modular design in terms of projection operators obtained from a design Density Matrix and shows their equivalence to the Linear Software Models results of the Laplacian matrix spectrum for the classical case. The application in practice of the design procedure for both classical and quantum software is illustrated by case studies.

## I. INTRODUCTION

Linear Software Models [7] represent classical software systems by a ***bipartite graph*** with two sets of vertices, one set standing for **Structors** – a generalization of *classes* – and another for **Functionals** – a generalization of class *methods* in object-oriented parlance. Structors contain and provide Functionals. Being a bipartite graph [18], there are edges only between vertices of the Structors set and vertices of the Functionals set, but not between vertices of the same set.

The Laplacian Matrix [13] [19] ***L*** associated with the bipartite graph is defined by eq. (1):

$$L = D - A \qquad (1)$$

where ***D*** is the Degree matrix – diagonal by definition – showing bipartite graph vertex degrees, and ***A*** an Adjacency matrix showing vertex neighbors. When two vertices are neighbors, the respective Adjacency matrix element is 1-valued, with a minus sign due to eq. (1). Otherwise, it is zero-valued.

### A. Laplacian for Classical Software Systems

Modules of a classical software system can be formally obtained by a procedure relying upon the eigenvalues and eigenvectors of the respective Laplacian. A module is defined as a connected component of the bipartite graph.

The number of modules of the software system represented by the Laplacian is given by the multiplicity of the zero-valued eigenvalues [9], [5]. The modules composition – in terms of Structors and Functionals – is given by the eigenvectors corresponding to the zero-valued eigenvalues.

When there are "outliers" – seen as Laplacian matrix elements coupling two potential modules – leading to a larger sparse module, one can split this larger sparse module using the Fiedler eigenvector [9]. This Fiedler vector fits to the smallest non-zero eigenvalue of the Laplacian matrix.

A modular bipartite graph of a schematic abstract example of a software system is seen in Fig. 1. The corresponding Laplacian matrix is shown in Fig. 2. Modules are seen as diagonal blocks of the Adjacency matrix within the Laplacian. Laplacian eigenvectors corresponding to the modules, are shown in Fig. 3.

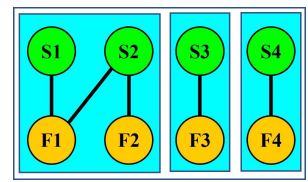

Fig 1. Bipartite Graph of a schematic abstract software system. It has 4 Structors – S1, S2, S3, S4 (green) and 4 Functionals – F1, F2, F3, F4 (orange). The Structor S2 provides two Functionals F1 and F2. This system has 3 modules (light blue rectangles) calculated from the Laplacian (in Fig. 2) through the eigenvectors (in Fig. 3). (Figures in color online).

|    | F1 | F2 | F3 | F4 | S1 | S2 | S3 | S4 |
|----|----|----|----|----|----|----|----|----|
| F1 | 2  | 0  | 0  | 0  | −1 | −1 | 0  | 0  |
| F2 | 0  | 1  | 0  | 0  | 0  | −1 | 0  | 0  |
| F3 | 0  | 0  | 1  | 0  | 0  | 0  | −1 | 0  |
| F4 | 0  | 0  | 0  | 1  | 0  | 0  | 0  | −1 |
| S1 | −1 | 0  | 0  | 0  | 1  | 0  | 0  | 0  |
| S2 | −1 | −1 | 0  | 0  | 0  | 2  | 0  | 0  |
| S3 | 0  | 0  | −1 | 0  | 0  | 0  | 1  | 0  |
| S4 | 0  | 0  | 0  | −1 | 0  | 0  | 0  | 1  |

Fig. 2. Laplacian matrix of the schematic abstract software system (in Fig. 1). The Laplacian diagonal contains the Degree matrix elements (pink). The Adjacency matrix (with minus signs) and its 3 block-diagonal modules (light blue rectangles) is seen in the upper-right and lower-left quadrants. Modules are obtained from the Laplacian eigenvectors (in Fig. 3).

| Eigen Vectors | F1 F2 F3 F4 S1 S2 S3 S4 | Modules |
|---|---|---|
| #1 | (1, 1, 0, 0, 1, 1, 0, 0) | {F1, F2, S1, S2} |
| #2 | (0, 0, 1, 0, 0, 0, 1, 0) | {F3, S3} |
| #3 | (0, 0, 0, 1, 0, 0, 0, 1) | {F4, S4} |

Fig. 3. Eigenvectors of the Laplacian matrix (in Fig. 2). The upper row (yellow) shows vertices fitting the eigenvector elements. The lower rows show three eigenvectors fitting zero-valued eigenvalues. The inferred modules, one module for each eigenvector, are shown in the r.h.s. (light blue).

### B. The Density Matrix Design Choice

The *Quantum Software Models* introduced in this paper were inspired by *Linear Software Models* and are their natural continuation due to their common basis upon linear algebra:

- Linear Software Models are for classical software systems design a framework based upon linear algebra;
- Quantum Software Models for quantum computing (e.g. [14]) also have linear algebra as its mathematical basis.

In the density operator picture of the Hilbert formulation of quantum mechanics the state of a quantum system is identified with a positive semidefinite, trace one, Hermitian matrix, called a Density Matrix [17]. The Laplacian matrix of a graph is symmetric and positive semidefinite. Braunstein and co-authors [3] observed that any Laplacian matrix $L(G)$ of a graph G, scaled by the degree-sum $d(G)$ of the graph G, has trace one, thus it is a Density Matrix $\rho$. They define it as the density matrix of a graph G:

$$\rho = L(G) / d(G) \qquad (2)$$

Since we represent any given software system by a bipartite graph, and its Laplacian, we re-define the density matrix in equation (2) as the ***design* Density Matrix** of the Software System.

We select the *design* Density Matrix as the starting point of choice for the design procedure of Quantum Software Models, by generality considerations, to make concrete the claim that there should be just a single unified and rigorous design procedure for both classical and quantum software systems:

- Any software system can be designed from the information in the *design* Density Matrix.

Except for the *design* qualifier, we look at the *design* Density Matrix as a density matrix for all purposes. We shall not explicitly use the *design* qualifier, unless needed to stress this characteristic. Now, we may focus on modularity from a deeper perspective.

### C. Paper Organization

The remaining of the paper is organized as follows. Section II looks at modular design from a deeper perspective. Section III formulates and illustrates Classical Software Design in terms of the Density Matrix. Section IV formulates and illustrates Quantum Software Design from the same perspective. Section V mentions related work. The paper is concluded with a Discussion in section VI.

## II. A Deeper Perspective on Modularity Design

A Density Matrix is a matrix representation of a Density Operator, which is a projection operator. Using the Dirac bra-ket notation, a Density Operator $\rho$ is a general kind of ket-bra:

$$\rho = |\psi\rangle\langle\psi| \qquad (3)$$

where $\psi$ is a generic notation for a quantum state [14].

Any projection operator, in short, a projector, actually projects its argument into a sub-space of the relevant Hilbert state space. From this point of view, a module – previously defined as a connected component of the bipartite graph – is redefined in terms of projectors, obtaining modules from the *design* Density Matrix.

### A. Modules Defined by Projection Operators

The Density operator $\rho$ acts on the state space of the system. A set of orthonormal basis vectors, a set of kets, spans the state space. One can associate a projection operator with each of the kets in the basis set.

One assumes that a whole software system design, classical or quantum, is completely described by its density operator. The respective Density Matrix can be expressed as a sum of the projection operators of the kets in the basis set, with suitable coefficients. For instance, for the Density Matrix obtained by scaling the Laplacian in Fig. 2, the computational basis set is $|000\rangle$, $|001\rangle$, $|010\rangle$, $|011\rangle$, $|100\rangle$, $|101\rangle$, $|110\rangle$ and $|111\rangle$.

Modules are partitions of the whole software system, with internal interactions, spanning a sub-space of the whole software system. More formally, one can state the definition as follows.

**Definition 1: Module of a Software System.** A module of a software system is a sub-system of a given software system. It spans a sub-space of the space state of the whole software system, given by a partition of the projection operators of the kets in the system basis set, such that each module sub-space is orthogonal to all other module sub-spaces of the software system.

Case studies illustrating this definition are provided for classical and quantum software systems in the next sections.

### B. From Density Matrix to Modules

A procedure to obtain Modules from the software system *design* Density Matrix is as follows:

1. Apply the Density Matrix to each ket in the basis set spanning the state space of the software system.
2. Obtain the projection operators for each ket in the basis set.
3. Express the Density Matrix as a sum of the projection operators obtained in the previous step.
4. Partition the sum of projection operators into disjoint sets of projection operators, each partition fitting a different module.
5. The number of modules in the software system is the number of disjoint sets of projection operators.
6. The composition of each module is given by the basis kets subset of the respective projection operators.

## C. Modules Validation

A final step to obtain the actual modules of a given software system is to check whether the modules obtained by the procedure in the previous sub-section are irreducible. A problem, already mentioned in sub-section 1.1, could be caused by an "outlier" coupling two smaller modules into a larger and sparser module. This larger module is reducible in principle to the smaller modules.

## III. CLASSICAL SOFTWARE DESIGN

The software design purpose is to enable software system analysis and development. Information sources for a classical software system depend on the software life cycle development stage: UML class diagram, a source code program, an executable code. This section formulates theorems on number and components of classical software modules and illustrates them by a case-study.

### A. From Class Diagram to Density Matrix

The idea is very simple: a- from a class diagram obtain the software system bipartite graph; b- from the graph generate the Laplacian matrix; c- scale the Laplacian by the degree-sum *d(G)* of the graph G, by eq. (2) to get the Density Matrix.

The information items extracted from the class diagram are: 1- class names; 2- methods provided by each class; 3- possible relationships between classes, in particular inheritance. Inheritance can be inferred, from the bipartite graph or from the Laplacian matrix, when two or more classes provide the same method.

Production of a software system design is not a one-pass action. Usually, one suggests an initial design, which is analyzed, and eventually improved. There could be a few cycles of this nature.

### B. Number and Components of Classical Software Modules

The modules number and components obtained from the Density Matrix (see section II.*B*) are stated in the next theorems.

**Theorem 1: Number of Classical Software Modules.** The number of modules in a classical software system represented by its *design* Density Matrix is given by the number of partition classes of the basis kets' projectors corresponding to the Density Matrix of the software system.

**Proof**:

By the Fiedler theorems [9], software modules are obtained from the Laplacian matrix eigenvectors, fitting the zero-valued eigenvalues. Thus, the proof consists in showing that projectors applied to the lowest Laplacian eigenvectors also obtain zero eigenvalues. It suffices to refer to the Laplacian, since it is related to the Density Matrix by eq. (2).

The lowest frequency Laplacian eigenvectors are non-negative and have two identical halves (by Theorem 4 in Exman and Sakhnini [7]), so *eigenvectors do not contribute opposite signs*.

Basis vectors (kets and bras) for the Density Matrix are mutually orthogonal. In the chosen computational basis each basis vector has a single positive element (different from all other basis vectors) and all other elements are zeros. Each partition *projector* is composed of pairs of *basis vectors*, characterizing the row and column of each matrix element. Applying the projector *bras* on the eigenvector *ket* obtains a zero eigenvalue due to different location of its non-zero elements and the *opposite bra signs*. □

**Theorem 2: Components of a Classical Software Module.** The module components in a classical software system represented by its Density Matrix are given by the Structors and Functionals fitting the respective basis kets/bras in the projection operators of the partition class of the software system Density Matrix.

**Proof**:

Since the partition classes of the projection operators are also partition classes of the kets/bras in the Density Matrix, and there is a one-to-one correspondence with the respective Structors and Functionals, the theorem is proved. □

### C. Classical Case Study: Prototype Design Pattern

Our classical software system case study has been shown in abstract form in Figures 1 to 3 (in section I.*A*). We continue with the same system, revealing that it is the *Prototype* design pattern. A design pattern is a reusable small set of classes with a definite role, frequently found in object-oriented programs.

The Prototype design pattern (see page 117 in the GoF book [11]) creates new objects by copying a prototypical instance. This system starts with the *Main* program of the <u>Prototype-Client</u> demanding a <u>Specific-Shape</u> (e.g. Rectangle, Triangle or Circle). If the shape is already stored in the <u>Shapes-Cache</u>, one *retrieves* the desired shape. Otherwise, one clones the desired shape, (the *Clone* functional is inherited from the <u>Generic-Cloneable-Shape</u>). A commercial Java code of the Prototype pattern similar to our model is found in ref. [16]. The Structors and Functionals of the Prototype design pattern are collected in Fig. 4. Its Density Matrix is seen in Fig. 5.

After collecting all the projection operators composing the Density Matrix and partitioning them into disjoint sets, one obtains the results in Fig. 6.

| Structors | | Functionals | |
|---|---|---|---|
| S1 | Generic-Cloneable-Shape | F1 | Clone |
| S2 | Specific-Shape | F2 | Calc-specific-Shape |
| S3 | Shapes-Cache | F3 | Load/Retrieve-Cache |
| S4 | Prototype-Client | F4 | Main |

Fig. 4. Prototype Design Pattern – List of Structors and Functionals, corresponding to the Laplacian matrix (in Fig. 2), and to the Density Matrix (in Fig. 5).

|       |     | \|000⟩ | \|001⟩ | \|010⟩ | \|011⟩ | \|100⟩ | \|101⟩ | \|110⟩ | \|111⟩ |
|---|---|---|---|---|---|---|---|---|---|
|       |     | F1 | F2 | F3 | F4 | S1 | S2 | S3 | S4 |
| ⟨000\| | F1 | 2  | 0  | 0  | 0  | −1 | −1 | 0  | 0  |
| ⟨001\| | F2 | 0  | 1  | 0  | 0  | 0  | −1 | 0  | 0  |
| ⟨010\| | F3 | 0  | 0  | 1  | 0  | 0  | 0  | −1 | 0  |
| ⟨011\| | F4 | 0  | 0  | 0  | 1  | 0  | 0  | 0  | −1 |
| ⟨100\| | S1 | −1 | 0  | 0  | 0  | 1  | 0  | 0  | 0  |
| ⟨101\| | S2 | −1 | −1 | 0  | 0  | 0  | 2  | 0  | 0  |
| ⟨110\| | S3 | 0  | 0  | −1 | 0  | 0  | 0  | 1  | 0  |
| ⟨111\| | S4 | 0  | 0  | 0  | −1 | 0  | 0  | 0  | 1  |

$\rho = 0.1*$

Fig. 5. Prototype Design Pattern – Density Matrix $\rho$ fitting the Laplacian matrix *L* (in Fig. 2). The Laplacian Trace (degree-sum) equals 10, therefore $\rho = 0.1*L$ by eq. (2). The basis set kets are shown above the respective columns (orange) and the fitting basis set bras to the left of the respective rows.

| Module Projectors | | Modules |
|---|---|---|
| #1 | (\|000⟩ - \|100⟩)(⟨000\| - ⟨100\|) + (\|000⟩ - \|101⟩)(⟨000\| - ⟨101\|) + (\|001⟩ - \|101⟩)(⟨001\| - ⟨101\|) | {F1, F2, S1, S2} |
| #2 | (\|110⟩ - \|010⟩)(⟨110\| - ⟨010\|) | {F3, S3} |
| #3 | (\|011⟩ - \|111⟩)(⟨011\| - ⟨111\|) | {F4, S4} |

Fig. 6. Prototype Module Projectors – Results obtained by applying the procedure in sub-section II.*B*. The projectors in terms of kets and bras are shown in the middle of the figure (omitting the degree-sum coefficient to emphasize the kets/bras partition classes). The module components in terms of Structors and Functionals are seen in the r.h.s. of the figure.

One should not confuse the ket above each column in Fig. 5 with the respective column label. To obtain the projectors fitting each ket one needs to apply the Density Matrix on the ket, obtaining the column labelled by the respective Structor label. For instance, applying $\rho$ |000⟩ one obtains the leftmost column labelled F1. Thus, the resulting projection operator is: $\rho$ |000⟩ = 0.1*(2*|000⟩ - |100⟩ - |101⟩)⟨000|.

The results in Fig. 6 comply with Theorems 1 and 2 in section III.*B*. The number of modules equals the number of projector partitions. The module components are given by the Structor and Functional labels, fitting the kets and bras within the projectors. They are confirmed by the classical linear algebra results from the eigenvectors (compare with Fig. 3).

## IV. QUANTUM SOFTWARE DESIGN

This section introduces Quantum Software Design from a new viewpoint on Quantum Computing. A single unified and rigorous design procedure for both classical and quantum software systems implies analogous techniques and the same theorems of the classical case (section III.*B*) to obtain quantum software modules. High-level quantum circuits are the source of quantum software design information.

### A. From High-Level Quantum Circuit to Density Matrix

First, one informally defines a high-level Quantum Circuit. It has parallel horizontal qubit lines and boxes containing one or more quantum gates (e.g. CNOT, Hadamard, Toffoli) or even classical computations. Boxes cover one or more qubit lines. A high-level quantum circuit is a sequential diagram, with "time" increasing from left (the input qubits) to right (typically a measurement output). There may be boxes (displayed vertically) executed in parallel. For more formal quantum circuit definitions see e.g. [14].

Information extraction is done as follows: a- begin with a high-level quantum circuit; b- extract lists of Structors and Functionals, and their relationships; c- obtain a bipartite graph from these concepts; d- generate the graph's Laplacian; e- obtain the fitting quantum software *design* Density Matrix.

### B. Analogies between Class Diagram and Quantum Circuit

High-level quantum circuits for quantum software design clearly have differences from class diagrams for classical design (see the Discussion section VI.*D*). Here we focus on similarities relevant to software design. Both class diagrams and quantum circuits expose software structures without fixing their exact numbers. For instance, in the classical Prototype case study, the number of Specific Shape classes is not fixed a priori. The same is true in a high-level quantum circuit: e.g. the number of applied Hadamard gates is left indeterminate.

### C. Structors and Functionals from Quantum Circuits

The data extracted from high-level quantum circuits is similar to the classical case: a set of Structors, a set of Functionals and their relationships, i.e. which Structor provides certain Functionals. These entities yield a bipartite graph and its quantum software *design* Density Matrix.

Structors and Functionals of a quantum software system design have the same roles as those entities in a classical software design. Structors – structural entities, the Boxes – are the basic building blocks of the hierarchical software structure. They are analogous to the *boxed subcircuits* of e.g. the Quipper quantum programming language [12]. These subcircuits are used multiple times within a larger circuit, thus boxed and given a generic name, semantically meaningful for the software engineer. Functionals – behavioral entities – are sets of gates for well-defined computations.

Modules, obtained from a quantum software design Density Matrix, enclose Structors and their Functionals. Modules containing modules, build the software system overall hierarchical structure.

### D. Quantum Case Study: Grover Search

Grover search is a well-known quantum algorithm for searching an unstructured database, attaining a quadratic speedup on the number of queries, relative to the classical computation. This quantum algorithm (see e.g. Nielsen and Chuang [14]) starts with equal probabilities for all input qubits, then recognizes and marks the target by an oracle, to iteratively amplify it in every cycle, and finally obtain the target by a measurement action. This is done in four steps, seen in the high-level quantum circuit in Fig. 7:

1. The $n^{th}$ tensor power of the <u>Hadamard operator **H**</u> transforms the input into an equal <u>superposition</u> state.

2. Apply an <u>oracle</u> to recognize and mark the target.

3. Perform target <u>amplification</u> by means of an "<u>inversion about the average</u>".

4. <u>Measure</u> the amplified target, yielding the final result.

The respective Structors and Functionals extracted from the high-level quantum circuit in Fig. 7 are shown in Fig. 8.

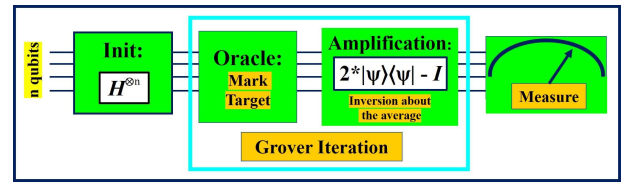

Fig. 7. Schematic Grover algorithm high-level quantum circuit – From left to right, the $n^{th}$ tensor power of the Hadamard operator *H* puts n input qubits in an equal superposition state. An oracle marks the target, which is amplified by an inversion about the average. The final result is obtained by measurement.

| Structors | | Functionals | |
|---|---|---|---|
| S1 | Init | F1 | Equal-Superposition |
| S2 | Oracle | F2 | Mark-Target |
| S3 | Amplification | F3 | Inversion about the average |
| S4 | Measurement | F4 | Measure |

Fig. 8. Grover algorithm Structors and Functionals – These correspond to the four boxes of the quantum circuit in Fig. 7.

The Oracle and the Amplification Structors, highlighted (in blue) in Figures 7 and 8, jointly constitute the Grover Iteration, looping in a few cycles during computation. A design justification for keeping these two Structors together in a single module is their functional similarity – analogous to a classical inheritance: Amplification is written as *2\*|ψ⟩⟨ψ| -I*; the Oracle can be formulated as *I- 2\*|ω⟩⟨ω|* (see e.g. Arikan et al. [1])**.** Here ω stands for the marked target.

The joint Grover Iteration is a size 2*2 software Module, in the middle of the quantum circuit (Fig. 7). This module relative position is sequential information relevant to a sequence diagram (see Discussion in section VI.*D*).

The Grover algorithm bipartite graph is an architectural units' diagram, generating a system *design* Density Matrix. One can easily perceive that, except for the above mentioned module sequential relative position, the Grover algorithm Density Matrix is almost identical to the classical Prototype Design Pattern Density Matrix.

An alternative architectural design of the Grover system keeps the Oracle and the Amplification separate, allowing independent optimization of each of them (e.g. Figgatt et al. [10]). Such *design* Density Matrix has Adjacency matrix quadrants with strictly diagonal modules.

## V. RELATED WORK

### A. Graphs, Laplacians and Density Matrices

Linear Software Models for classical software system design based upon linear algebra have been developed by Exman and co-authors. Exman and Sakhnini described software systems by bipartite graphs, leading to Laplacian Matrices [7]. Splitting too sparse software modules has been done with Fiedler eigenvectors [9], [5]. Exman and Wallach [8] recently applied these Models to software consumers.

Braunstein and co-authors [3], followed by Wu [20], make the transition from graphs' Laplacian matrices to quantum computing Density Matrices, investigating separability issues.

Perez-Delgado and Perez-Gonzalez [15], in a non-algebraic approach to Quantum Software Modeling, suggest minimal quantum extensions to the UML *classical language*, in order to apply it to quantum software systems. This in contrast to our opposite direction, viz. Density Matrix *quantum language* to be applied to classical software systems.

### B. Modularity in Quantum Software Design

Modularity ideas for quantum computing software, have been recognized within several contexts. Zhang et al. [21] applied modular computer architecture to NMR quantum computing, claiming that modularized software architecture plays an increasing role for large-scale quantum computing.

Debnath et al. [6] demonstrate quantum computing programmable in software, compiled into modular logic gates for reconfigurable algorithms without altering the hardware.

Figgatt et al. [10] describe a complete 3-Qubit Grover search, with various Oracle implementations. The initialization and amplification stages were optimized disregarding the oracle contents to preserve the algorithm modularity, enabling insertion of possible alternative oracles without changing the other stages.

## VI. DISCUSSION

### A. Modularity Reasons for Quantum Design

A single quantum software system may have Density Matrices for distinct purposes, among others, based upon different choices of basis vectors. For instance, Batle et al. [2] use a Bell basis' Density Matrix, to investigate how quantum correlations vary as the Grover search algorithm is run.

The current paper chooses a *design* Density Matrix to analyze modularity of quantum software systems. We observe that any scaled Laplacian of a software system is a design Density Matrix, but not any density matrix in general can be converted to a Laplacian representing a software system design.

There are various reasons for quantum software modularization. These include enabling comprehension of quantum computation semantics by human engineers, and increasing computation efficiency, in particular partitioning of networked quantum systems. Often functional separation facilitates comprehension or enhances independent optimization. In other cases, the opposite may be needed, i.e. integrating various Functionals into a single Module.

### B. Unified Classical and Quantum Design Procedure

There is a double motivation for "a single unified and rigorous design procedure for both classical and quantum software systems" focusing on modularity. The pragmatic argument is to facilitate development of hybrid software systems made of classical and quantum sub-systems.

A foundational argument is to preserve Brooks' idea of *conceptual integrity* [4] throughout software systems involving both classical and quantum aspects, enabling comprehension of these systems by human software engineers.

### C. Classical Software Systems as classical limit of Quantum Systems

Counterintuitively at first sight, we conjecture that it should be easier to obtain classical software systems as a classical limit of quantum software systems, than the other way round. Indeed, in physics there is a theoretical expectation of classical systems to be derivable as classical limits from quantum systems. Moreover, this continuity between quantum and classical software systems, offers novel yet unexplored territory for deeper understanding of classical software (see next section VI.*D*).

Can we heuristically justify a quantum to classical software continuity? In one sentence the argument is: the quantum state/operator duality is a suitable formalism for the same classical software state/operator duality.

In the density operator picture of quantum mechanics a quantum system *state* is identified with the Density Matrix, which at the same time is an *operator* applicable to states. Von Neumann's Density Matrix [17] insight is supported by elegant Dirac notation. A ket |a⟩ (and a bra ⟨b|) is a *state*. A bra-ket inner product ⟨b|a⟩ yields a number. By simple order exchange, |a⟩⟨b| is an *operator*, a projector, applicable to other states.

Classical software is a static description of a system, a potential computation waiting for a trigger, in other words, a *state*. At the same time, classical software is runnable – when interpreted or compiled, details being irrelevant for the argument – i.e. an *operator* applicable to other input states.

*D. Software and Hardware: Design and Implementation*

The original design diagrams for classical and quantum software are different mainly by historical reasons. Classical design diagrams include: UML class diagram displaying structure; sequence diagram showing time-dependence of specific scenarios that may occur in a system; statechart displaying states for the whole system.

Quantum design diagrams include high-level quantum circuits, displaying a double character of both structure and time sequence, with implicit states. In this paper we focus on structure design, leaving the sequential aspect to be discussed elsewhere.

In this work *Quantum Software design* means high-level abstract Structors, such as init or oracle, and their Functionals, as illustrated by the Grover algorithm case study. Quantum software *implementation* makes design concrete by assigning to abstract design entities, specific types and numbers of quantum gates, interconnections, input and output qubits.

*Quantum hardware* design and implementation refer to the underlying physics – e.g. NMR, trapped atomic ions, photonic devices – easily distinguishable from quantum software, and out of the scope of this paper.

*E. Future Work*

This paper made initial steps towards a unified modeling viewpoint of classical and quantum software systems. In future work we intend to explore additional theoretical aspects of the ideas presented here and apply them in practice to a more extensive investigation of medium and larger quantum software systems.

*F. Main Contribution*

The main contribution of this paper is the unified linear algebra design approach to classical and quantum software systems, starting from their representation as *design* Density Matrices.